\documentclass[numberedappendix]{emulateapj}

\usepackage{epsfig, natbib}
\tightenlines

\countdef\decade=200
\advance\decade by \year
\countdef\hours=201	
\advance\hours by \time
\divide\hours by 60
\countdef\mins=202
\advance\mins by \hours
\multiply\mins by 60
\multiply\hours by 100
\countdef\miltime=203
\advance\miltime by \hours
\advance\miltime by \time
\advance\miltime by -\mins

\newcommand{\ltsim}{\protect\raisebox{-0.5ex}{$\:\stackrel{\textstyle <}
        {\sim}\:$}}
\newcommand{\gtsim}{\protect\raisebox{-0.5ex}{$\:\stackrel{\textstyle >}
        {\sim}\:$}}

\newcommand{\tff}{t_{\rm ff}}
\newcommand{\sfrff}{\mbox{SFR}_{\rm ff}}
\newcommand{\fmol}{f_{\rm H_2}}

\newcommand{\rhomol}{\rho_{\rm H_2}}
\newcommand{\Sigmag}{\Sigma_{\rm g}}

\newcommand{\Sigmasfr}{\dot{\Sigma}_*}
\newcommand{\msun}{M_{\odot}}
\newcommand{\hi}{H~\textsc{i}}

\newcommand{\Sigmacl}{\Sigma_{\rm cl}}

\newcommand{\pcl}{P_{\rm cl}}

\newcommand{\avir}{\alpha_{\rm vir}}

\newcommand{\kb}{k_{\rm B}}

\newcommand{\calm}{\mathcal{M}}

\defcitealias{krumholz08c}{KMT08}
\defcitealias{krumholz08d}{KMT09}
\defcitealias{krumholz05c}{KM05}
\defcitealias{bigiel08a}{B08}

\begin{document}

\title{The Star Formation Law in Atomic and Molecular Gas}


\author{Mark R. Krumholz}
\affil{Department of Astronomy \& Astrophysics, University of California, Santa Cruz, CA 95060}
\email{krumholz@ucolick.org}

\author{Christopher F. McKee}
\affil{Departments of Physics and Astronomy, University of California, Berkeley, CA 94720-7304}
\email{cmckee@astro.berkeley.edu}

\author{Jason Tumlinson}
\affil{Space Telescope Science Institute, Baltimore, MD 21218}
\email{tumlinson@stsci.edu}

\begin{abstract}
We propose a simple theoretical model for star formation in which the local star formation rate in a galaxy is determined by three factors. First, the interplay between the interstellar radiation field and molecular self-shielding determines what fraction of the gas is in molecular form and thus eligible to form stars. Second, internal feedback determines the properties of the molecular clouds that form, which are nearly independent of galaxy properties until the galactic ISM pressure becomes comparable to the internal GMC pressure. Above this limit, galactic ISM pressure determines molecular gas properties. Third, the turbulence driven by feedback processes in GMCs makes star formation slow, allowing a small fraction of the gas to be converted to stars per free-fall time within the molecular clouds. We combine analytic estimates for each of these steps to formulate a single star formation law, and show that the predicted correlation between star formation rate, metallicity, and surface densities of atomic, molecular, and total gas agree well with observations.
\end{abstract}

\keywords{galaxies: ISM --- ISM: clouds --- ISM: molecules --- stars: formation}

\section{Introduction}
\label{intro}

The last decade has seen a revolution in our understanding of star formation in galaxies, driven by the advent of spatially resolved multi-wavelength surveys. Prior to this work, our observational constraints on the star formation process were largely limited to low resolution surveys that characterized entire galaxies using only a handful of observable quantities, e.g.\ the mean surface density of star formation averaged over a whole disk. While these surveys yielded a number of intriguing results -- most famously the \citet{kennicutt98a} star formation law -- they left unanswered many basic questions about the physics of star formation. For example, they could not clearly determine whether star formation correlates more strongly with the molecular or total gas content of a galaxy \citep{kennicutt98a, wong02}, or whether star formation is regulated primarily by local processes within individual star-forming clouds (e.g.\ \citealt{krumholz05c}, hereafter \citetalias{krumholz05c}; \citealt{shu07}) or by galactic-scale processes such as spiral shocks, supernovae, or cloud-cloud interactions \citep[e.g.][]{wyse86a, silk97, tan00, li05a}.

Now, however, emission maps at 24 $\mu$m from the {\it Spitzer} Infrared Nearby Galaxy Survey (SINGS) provide us with accurate estimates of the rate of dust-enshrouded star formation at resolutions of better than a kpc in nearby galactic disks \citep{kennicutt03, kennicutt07a, calzetti07}, while ultraviolet observations from the GALEX Nearby Galaxy Survey (NGS) reveal the rates of non-obscured star formation with comparable resolution and accuracy \citep{gil-de-paz07a}. Observations of \hi\ emission from the VLA as part of The \hi\ Nearby Galaxy Survey \citep[THINGS;][]{walter08a} and of CO emission by the BIMA Survey of Nearby Galaxies \citep[BIMA SONG;][]{helfer03} and the HERA CO-Line Extragalactic Survey \citep[HERACLES;][]{leroy09a} using the 30 m IRAM telescope provide maps of the gas content of galaxies at comparable resolutions. Combining these data sets leads to two particularly important conclusions that our theoretical models of star formation must incorporate and explain.

The first observational result is that that star formation is a direct product of the molecular gas in a galaxy, not of all the gas. Across a wide range of galactic environments the star formation rate correlates well with molecular gas, and poorly or not at all with the atomic gas measured on sub-kpc spatial scales (\citealt{wong02, kennicutt07a, leroy08a, bigiel08a}, hereafter \citetalias{bigiel08a}). The correlation between star formation rate and molecular gas surface density is super-linear in samples that extend to starbursts with gas surface densities $\sim 10^3$ $\msun$ pc$^{-2}$ \citep{kennicutt98a}, but is nearly linear for galaxies with molecular surface densities from $5-100$ $\msun$ pc$^{-2}$ (\citetalias{bigiel08a}; however, see \citealt{kennicutt07a} for a different interpretation). Observations of the low density outskirts of galactic disks hint that the linearity may break down there \citep{gardan07a, fumagalli08a}, but it is unclear if this indicates a change in the star formation process or a change in the CO to H$_2$ conversion factor.

The second observational result is that giant molecular clouds (GMCs) have remarkably similar properties in all nearby galaxies. Across the Local Group GMCs appear to have the same surface density, roughly $85$ $\msun$ pc$^{-2}$, and to obey the same linewidth-size relation \citep{blitz07a, bolatto08a, heyer08a}. Together these two observations imply that all observed molecular clouds are not far from virial balance between gravity and internal turbulence. There is of course considerable uncertainty in the GMC surface density, arising mostly from the CO-to-H$_2$ conversion factor, but it is striking that there is no clear evidence of a systematic trend in GMC properties across a sample of galaxies ranging from \hi-dominated dwarfs to molecule-rich giant spirals. This seems to be a critical clue to the physics of molecular clouds.

Efforts to incorporate these observational results and their implications into numerical simulations are already underway. \citet{robertson08a} and \citet{gnedin09a} present calculations of galaxy evolution that include models for the chemistry of hydrogen molecule formation and destruction, and that restrict star formation to occur only in molecular gas. \citet{tasker09a} use high resolution simulations of galactic disks to study the origin of the observed properties of GMCs. Our goal in this paper is to complement and extend this work by developing a simple analytic model for the star formation law that is based on the insights provided by the new observational surveys, and that incorporates the theoretical understanding that has developed around them.

Our basic approach is to break the problem into three tractable pieces, each of which can be treated using models and observations already available in the literature. The first is the problem of calculating what fraction of the gas in a given portion of a galaxy will be in the molecular phase and thus eligible to form stars. The problem of determining the molecular fraction in the ISM has been treated extensively \citep[e.g.][]{vandishoeck86, sternberg88, elmegreen93, draine96, browning03}; in our approach we will adopt the model of  \citet[hereafter KMT08 and KMT09]{krumholz08c, krumholz08d}, which agrees very well with the observations of molecular fractions reported by \citet{blitz04} and \citet{leroy08a}, and has the advantage that it does not depend on unknown and generally unmeasurable quantities such as the intensity of the ultraviolet radiation field inside a galaxy. The second problem is to estimate the characteristic properties of the GMCs in a galaxy, which can be done using a combination of their observed properties simple arguments based on virial balance and the galactic Jeans mass (e.g.\ \citealt{kim01}; \citetalias{krumholz05c}). The third problem is to estimate the rate at which molecular clouds of known properties transform themselves into stars. This rate is known from observations to be $\sim 1\%$ of the mass per free-fall time \citep{zuckerman74, krumholz07e}, a value that can be understood theoretically as a result of regulation of star formation by supersonic turbulence (\citealt{kravtsov03}; \citetalias{krumholz05c}; \citealt{wada07a}). In \S~\ref{sflawsec} we explain how these three components can be combined to produce a star formation law, and then in \S~\ref{obscomp} we compare the results to observations.

\section{The Star Formation Law}
\label{sflawsec}

Given that star formation occurs in molecular gas, we formulate our theoretical law for the local star formation rate (SFR) surface density $\Sigmasfr$ in a galaxy as a product of three factors:
\begin{equation}
\label{sfreq1}
\Sigmasfr = \Sigmag \fmol\frac{\sfrff}{\tff}.
\end{equation}
Here $\Sigmag$ is the total gas surface density at some point in the galaxy. In practice this will always be an average over some size scale, determined by the resolution of observations (or simulations). This determines the total available ``raw material" for star formation. The factor $\fmol$ is the fraction of this mass in molecular form; atomic gas does not participate in star formation. The molecular component of the gas is organized into clouds which have some mean volume density $\rhomol$, and $\tff=[3\pi/(32 G \rhomol)]^{1/2}$ is the free-fall time at this mean density. The quantity $\sfrff$ is the dimensionless star formation rate; it is the fraction of the gas transformed into stars per free-fall time. Alternately, one may think of it as the star formation efficiency over one free-fall time (as opposed to the total star formation efficiency, which might mean the fraction of gas transformed into stars over some other timescale, such as the galactic rotation time or the lifetime of an individual GMC). The third factor, $\sfrff/\tff$, is simply the SFR per free-fall time divided by the free-fall time, which is the inverse of the time required to convert all of the gas into stars. To make a model for the star formation law, we must estimate $\fmol$, $\tff$, and $\sfrff$ in terms of the observable quantities for a galaxy. 

\subsection{The Molecular Fraction}

The molecular mass fraction $\fmol$ is determined by the balance between dissociation of molecules by the far ultraviolet (FUV) interstellar radiation field in the Lyman-Werner bands and formation of molecules on the surfaces of dust grains. \citetalias{krumholz08c} and \citetalias{krumholz08d} show that to good approximation $\fmol$ within a single atomic-molecular complex is a function of the gas surface density of the complex $\Sigma_{\rm comp}$ and the metallicity $Z$. We will not repeat the full derivation of this result here, but a summary of the calculation that produces it is that one first solves the idealized problem of finding where the transition between the atomic envelope and the molecular interior occurs within a uniform sphere of hydrogen gas and dust embedded in an isotropic dissociating radiation field. This analysis shows that the fraction of the complex in molecular form depends on the dust optical depth of the complex $\Sigma_{\rm comp}\sigma_d$ and on the dimensionless ratio $\chi \propto \sigma_d G_0 / (n_{\rm CNM} \mathcal{R})$, where $\sigma_d$ is the dust cross section per hydrogen nucleus, $G_0$ is the intensity of the dissociating radiation field, $n_{\rm CNM}$ is the number density of gas in the cold atomic medium that surrounds the molecular part of the cloud, and $\mathcal{R}$ is the rate coefficient for H$_2$ formation on the surfaces of dust grains. Since $\sigma_d$ and $\mathcal{R}$ are both, to first order, simply measures of the total amount of dust in a galaxy, their ratio should not vary widely between galaxies. Similarly, in a galaxy with a two-phase atomic medium the cold atomic gas density $n_{\rm CNM}$ is determined by thermal pressure balance between the two phases, which in turn depends on the balance between heating by FUV photons and atomic line cooling in the atomic gas. Analysis of these processes implies that the ratio $G_0/n_{\rm CNM}$ is a weak function of metallicity and is otherwise independent of galaxy properties \citep{wolfire03}. Thus $\chi$ varies little between galaxies, and this result enables us to write the molecular fraction for a given atomic-molecular complex as a function solely of its gas surface density $\Sigma_{\rm comp}$ and its metallicity $Z$:
\begin{equation}
\label{fmolapprox}
\fmol(\Sigma_{\rm comp},Z') \approx 1 - \left[1+\left(\frac{3}{4}\frac{s}{1+\delta}\right)^{-5}\right]^{-1/5}
\end{equation}
where $s = \ln(1+0.6\chi)/(0.04\Sigma_{\rm comp,0} Z')$, $\chi = 0.77(1+3.1 Z'^{0.365})$, $\delta = 0.0712 (0.1 s^{-1} + 0.675)^{-2.8}$, $\Sigma_{\rm comp,0} = \Sigma_{\rm comp}/(1\,\msun\mbox{ pc}^{-2})$, and $Z'$ is the metallicity normalized to the solar value. Note that this approximation is slightly different with the one given in \citetalias{krumholz08d}; the two agree to within a few percent for clouds that are substantially molecular, but this one is more accurate at small molecular fractions (McKee, Krumholz, \& Tumlinson, 2009, in preparation).

Here $\Sigma_{\rm comp}$ is the surface density of a $\sim 100$ pc-sized atomic-molecular complex. However, extragalactic observations generally measure a gas surface density $\Sigmag$ that is averaged over a much larger scale. Since $\fmol$ increases super-linearly with $\Sigma_{\rm comp}$, clumping of the gas on scales below the observational resolution would lead us to underpredict $\fmol$ if we were simply to use the large-scale-averaged value of $\Sigmag$ in place of $\Sigma_{\rm comp}$ equation (\ref{fmolapprox}). Since we wish to propose a model that is applicable to data and simulations at a range of resolutions, it is convenient to approximately correct for this effect by letting $\Sigma_{\rm comp} = c \Sigmag$, where $c\ge 1$ is a clumping factor and $c\rightarrow 1$ as the resolution approaches $\sim 100$ pc.

As a final caveat, it is important to point that our calculation of $\fmol$ in \citetalias{krumholz08d} assumes that the \citet{wolfire03} semi-analytic model for the atomic ISM is applicable, and the model begins to break down at metallicities below roughly 5\% of solar (see Figure 13 of \citealt{wolfire03}) because dust grains and polycyclic aromatic hydrocarbons begin to be neutral rather than positively charged, as the model assumes. Turbulent heating of the cold \hi\ phase \citep{pan09a}, which is not included in the \citeauthor{wolfire03} models, is also likely to be important at low metallicity. Thus, although our general method of calculating molecular fractions will apply even at low metallicities, the relationship between $n_{\rm CNM}$ and $G_0$ which is used to derive equation (\ref{fmolapprox}) is not valid at metallicities $Z'<0.05$.

\subsection{Giant Molecular Cloud Properties}

Next we must compute $\tff$ and $\sfrff$, which will depend on the properties of the star-forming GMCs in a galaxy. Before proceeding with such a calculation, we note that observations of Local Group galaxies indicate that in galaxies ranging from metal-poor dwarfs to molecule-rich spirals, the molecular cloud surface density $\Sigmacl\approx 85$ $\msun$ pc$^{-2}$ and the molecular cloud virial ratio $\avir\approx 2$ independent of galactic environment \citep{blitz07a, bolatto08a}.\footnote{Note that this value of $\Sigmacl$ is lower than the $170$ $\msun$ pc$^{-2}$ for Galactic GMCs found by \citealt{solomon87}, but is consistent with the lower value determined by the more recent survey of \citealt{heyer08a}.} This invariance is reasonably easy to understand on theoretical grounds. The virial theorem implies that the mean pressure within the cloud is $\pcl/\kb = 0.7\times 10^5 \,\avir \Sigmacl'^2 \mbox{ K cm}^{-3}$ \citepalias{krumholz05c}, where $\Sigmacl'\equiv\Sigmacl/(85\,\msun\mbox{ pc}^{-2})$. In comparison, \citet{boulares90} find that the mean kinetic pressure\footnote{We consider only turbulent and thermal pressure because the mean galactic magnetic field and cosmic rays pervade GMCs and the  intercloud medium equally, and therefore provide neither support nor confining pressure.} in the ISM of a Milky Way-like galaxy is $1.4\times 10^4$ K cm$^{-3}$, an order of magnitude lower, although the pressure may be higher than average in spiral arms where GMCs form. The pressure is almost certainly lower in low surface-density dwarfs. The mismatch between the pressures in GMCs and the pressures in their environments indicates that external pressure is at most marginally important in determining the properties of molecular clouds, and that GMCs must instead be internally regulated; in effect, a GMC forgets about its galactic environment. Moreover, such a picture provides a quantitative explanation for the observed values of the GMC surface density and virial ratio. The dominant mechanism of internal regulation is H~\textsc{ii} region feedback \citep{matzner02}, a process whose efficiency depends on the column density of the cloud. \citet{krumholz06d} find that H~\textsc{ii} regions stabilize GMCs at column densities of roughly $100$ $\msun$ pc$^{-2}$ and virial ratios $\avir\approx 1-2$, consistent within the uncertainties with the observed values. 

The constant surface densities and virial ratios of GMCs provide a natural way to estimate $\tff$ and $\sfrff$. Consider a GMC of mass $M$, surface density $\Sigmacl$, and virial ratio $\avir$. The volume density in this cloud is $\rhomol \approx (3\pi^{1/2}/4) \Sigmacl^{3/2} M^{-1/2}$, so the free-fall time is
\begin{equation}
\label{tffeqn}
\tff = 8\, \Sigmacl'^{-3/4} M_6^{1/4} \mbox{ Myr}
\end{equation}
where $M_6=M/10^6\,\msun$. Similarly, \citetalias{krumholz05c} show that the star formation rate per free-fall time in a turbulent medium is approximately
\begin{equation}
\label{sfrffeqn}
\sfrff \approx 0.15 \epsilon_{\rm core} \avir^{-0.68} \calm^{-0.32},
\end{equation}
where $\calm$ is the 1-D Mach number of the turbulence and $\epsilon_{\rm core}$ is the fraction of the mass in a gravitationally-bound prestellar core that is incorporated into a star rather than being ejected by protostellar outflows. \citetalias{krumholz05c} adopt $\epsilon_{\rm core}=0.5$ based on analytic models showing $\epsilon_{\rm core} \approx 0.25-0.75$ \citep{matzner00}; more recent work suggests the true value is $\epsilon_{\rm core} \approx 0.3$ \citep{alves07}, so we adopt $\epsilon_{\rm core} = 0.3$. 

The virial ratio is related to the one-dimensional velocity dispersion $\sigma$ in a GMC by $\avir\equiv 5\pi^{-1/2}(M \Sigmacl)^{-1/2} \sigma^2/G$ \citep{bertoldi92}, so $\sigma = 3.7 \,\avir^{1/2} \Sigmacl'^{1/4} M_6^{1/4}$ km s$^{-1}$.\footnote{For our fiducial $\Sigmacl'=1$ and $\avir=2$, this agrees with the observed linewidth-size relation $\sigma\approx 0.44_{-0.13}^{+0.18} (R/\mbox{pc})^{0.60 \pm0.10}$ km s$^{-1}$ \citep{bolatto08a} to within the error bars.} For a molecular cloud temperature of 10 K the corresponding Mach number is $\calm=20 \avir^{1/2} \Sigmacl'^{1/4} M_6^{1/4}$, so 
\begin{equation}
\label{sfrffeqn1}
\sfrff\approx 0.017 \avir^{-0.84} \Sigmacl'^{-0.08} M_6^{-0.08}.
\end{equation}
This is consistent with the observed value $\sfrff\approx 0.01$ \citep{krumholz07e}.

Combining equations (\ref{tffeqn}) and (\ref{sfrffeqn1}), and adopting a fiducial value of $\avir=2$, gives
\begin{equation}
\label{sfrfftff}
\frac{\sfrff}{\tff} = \frac{\Sigmacl'^{0.67} M_6^{-0.33}}{0.8\mbox{ Gyr}}.
\end{equation}

The invariance of molecular cloud properties that we observe in nearby galaxies must break down in galaxies with sufficiently high surface densities, where the external pressure is no longer negligible compared to a GMC's internal pressure. Since pressure varies as $P\propto \Sigma^2$ for both molecular clouds and galactic disks \citepalias{krumholz05c}, the galactic environment will become significant in determining molecular cloud properties once the galactic surface density averaged over large scales becomes comparable to the surface density of an individual GMC. In this case the GMC surface density must increase in order to maintain pressure balance with the rest of the galaxy's ISM, which simply requires that $\Sigmacl\approx \Sigmag$ for $\Sigmag > 85$ $\msun$ pc$^{-2}$. (Alternately, \citealt{komugi06a} suggest that a change in GMC properties might be expected when $\Sigmag \sim 10^2-10^3$ $\msun$ pc$^{-2}$ because at such high surface densities collisions between GMCs become common.) Observations are consistent with this hypothesis: in the central kpc of M64, where the galactic surface density runs from $\sim 50-1000$ $\msun$ pc$^{-2}$, the GMC surface density is not constant, and instead rises with galactic pressure. Averaged over the entire galaxy the mean GMC surface density is $250$ $\msun$ pc$^{-2}$ \citep{rosolowsky05a}. Thus, the free-fall time in GMCs in high surface density galaxies varies as $\Sigmag^{-3/4}$. If we adopt a column density of $\Sigmacl=85$ $\msun$ pc$^{-2}$ for all GMCs in normal surface density galaxies and $\Sigmacl=\Sigmag$ at higher galactic surface densities, then we have
\begin{equation}
\label{sfrfftff1}
\frac{\sfrff}{\tff} = \frac{M_6^{-0.33}}{0.8\mbox{ Gyr}} \max\left[1,\left(\frac{\Sigmag}{85\,\msun\mbox{ pc}^{-2}}\right)^{0.67}\right].
\end{equation}

Equation (\ref{sfrfftff1}) gives an estimate for $\sfrff/\tff$ in a molecular cloud of a known mass. To complete the calculation, we must estimate the characteristic molecular cloud mass in a galaxy. We follow \citetalias{krumholz05c} in estimating that this will be determined by the Jeans mass in the galaxy, which is
\begin{equation}
\label{mcleq}
M \approx \frac{\sigma_{\rm g}^4}{G^2\Sigmag} = \frac{\pi^4 G^2 \Sigmag^3 Q^4}{4\Omega^4},
\end{equation}
where $\sigma_{\rm g}$ is the gas velocity dispersion, $Q$ is the Toomre $Q$ of the galactic disk, and $\Omega$ is the angular velocity of its rotation. If we can directly measure $\Sigmag$, $\Omega$, and $Q$, or $\Sigmag$ and $\sigma_{\rm g}$, for a galaxy, then we can solve for $M$ directly and substitute into equation (\ref{sfrfftff1}) to obtain a characteristic value of $\sfrff/\tff$ for that galaxy. However, often one or more of the quantities are unknown, and even when they are known it is useful to have a rough estimate in terms of a single quantity such as $\Sigmag$ rather than three quantities $\Sigmag$, $\Omega$, and $Q$. Since $M_6$ enters the star formation rate only to the $0.33$ power, any errors we make in this approximation are unlikely to have strong effects. We therefore follow \citetalias{krumholz05c} in assuming that all galaxies will be marginally Toomre stable, $Q\approx 1$, and noting that there is broad statistical correlation $\Omega/\mbox{Myr}^{-1} \approx  0.054 (\Sigmag/85\,\msun\mbox{ pc}^{-2})^{0.49}$. If we use this correlation in (\ref{mcleq}) then we obtain
\begin{equation}
\label{mcleq1}
M_6 \approx 37 \left(\frac{\Sigmag}{85\,\msun\mbox{ pc}^{-2}}\right)^{1.0}.
\end{equation}

Finally, it is worth noting here that our estimate of the molecular cloud volume density, which depends on $\Sigma_{\rm cl}$ and $M_6$, is somewhat different than that of \citetalias{krumholz05c}. They assumed that GMC surface densities were set largely by external pressure in a galaxy, and computed the density based on this assumption. As discussed above, more recent observational and theoretical work suggests that instead GMC densities are primarily set by internal feedback processes and do not vary significantly with galactic conditions, at least in Milky Way-like galaxies. Our model in this paper takes this result into account.

\subsection{The Full Star Formation Law}

We have now derived the major components of our star formation law (equation \ref{sfreq1}). The molecular fraction $\fmol$ depends only on gas surface density $\Sigmag$, metallicity $Z'$, and the clumping of the gas $c$ on scales unresolved in a given observation or simulation (equation \ref{fmolapprox}). It increases with $\Sigmag$, becoming fully molecular at $\sim 10/cZ'$ $\msun$ pc$^{-2}$. We have also derived an analytic relation for the inverse star formation timescale $\sfrff/\tff$ in two regimes. Where internal GMC pressure far exceeds the ambient ISM gas pressure and GMCs ``forget" their environment -- as typically occurs in nearby galaxies with $\Sigmag < 85$ $\msun$ pc$^{-2}$ --  this timescale does not depend on $\Sigmag$ except indirectly through the molecular cloud mass (equation \ref{mcleq1}). Above $\Sigmag = 85$ $\msun$ pc$^{-2}$, ambient pressure becomes comparable to the GMC internal pressure and the star formation timescale depends on $\Sigmag$ (equation \ref{sfrfftff1}). In neither case does the timescale depend on either the metallicity or the clumping, so the star formation rate in molecular gas does not depend on either of these quantities. Only the star formation rate in total gas does.

We are now ready to combine these pieces into our single star formation law:
\begin{eqnarray}
\dot{\Sigma}_* & = & f_{\rm H_2}(\Sigmag, c, Z') \frac{\Sigmag}{2.6\mbox{ Gyr}}
\nonumber \\
& & {}\times
\left\{
\begin{array}{ll}
\left(\frac{\Sigmag}{85\,\msun\,{\rm pc}^{-2}}\right)^{-0.33}, \quad & \frac{\Sigmag}{85\,\msun\,{\rm pc}^{-2}} < 1 \\
\left(\frac{\Sigmag}{85\,\msun\,{\rm pc}^{-2}}\right)^{0.33}, \quad & \frac{\Sigmag}{85\,\msun\,{\rm pc}^{-2}} > 1 
\end{array}
\right..
\label{sfreq2}
\end{eqnarray}

\begin{figure*}
\plotone{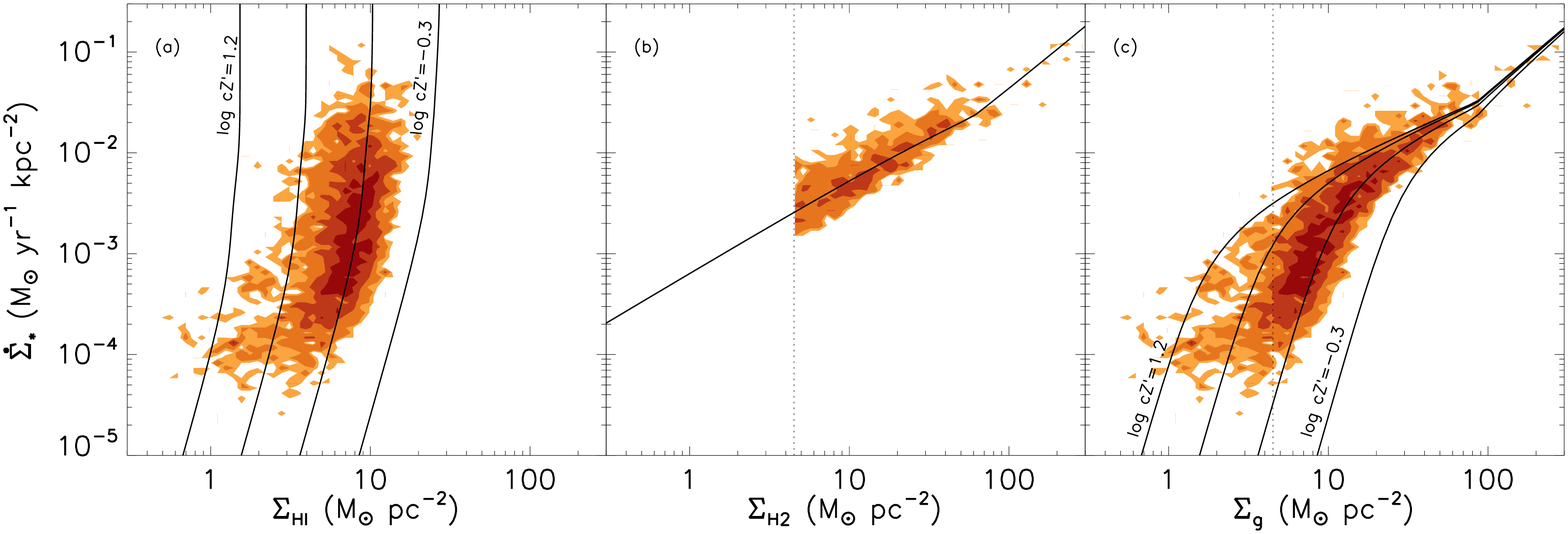}
\caption{
\label{sflaw1}
Star formation rate surface density $\dot{\Sigma}_*$ as a function of \hi\ ({\it panel a}), H$_2$ ({\it panel b}), and total gas ({\it panel c}) surface densities $\Sigma_{\rm HI}$, $\Sigma_{\rm H_2}$, and $\Sigma_{\rm g}$. Lines show our theoretical model predictions for values of clumping factor times metallicity of $\log cZ'=-0.3, 0.2, 0.7$, and $1.2$, as indicated. Contours show observations from THINGS,  and are constructed as in \citetalias{bigiel08a}: we break the plane of the plot into bins 0.05 dex wide in each direction and count the number of independent data points in each bin. The contours represent, from lightest to darkest, 1, 2, 5, and 10 data points. The dashed vertical lines in the $\Sigma_{\rm H_2}$ and $\Sigma_{\rm g}$ plots indicate the THINGS CO sensitivity limit of 4.5 $\msun$ pc$^{-2}$. Note that our plots are shifted by a factor of 1.36 relative to those of \citetalias{bigiel08a} because we include the mass of helium in $\Sigma_{\rm HI}$, $\Sigma_{\rm H_2}$, and $\Sigma_{\rm g}$.
}
\end{figure*}

\section{Comparison to Observations}
\label{obscomp}

\begin{figure*}
\plotone{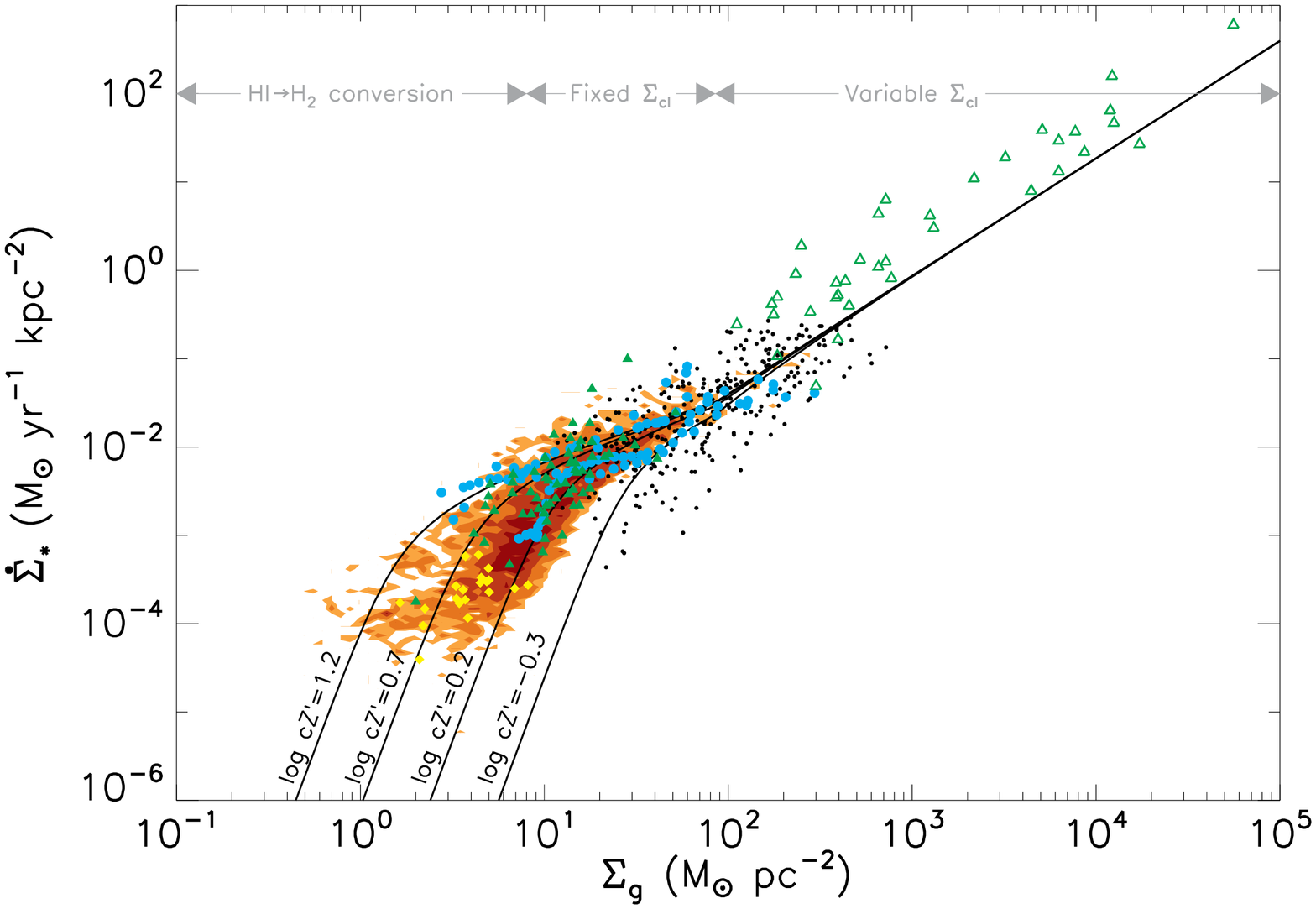}
\caption{
\label{sflaw2}
Star formation rate surface density $\dot{\Sigma}_*$ as a function of total gas surface density $\Sigma_{\rm g}$. Lines and contours are the same as in Figure \ref{sflaw1}. Other points are a compilation of literature data from \citetalias{bigiel08a}. We show individual apertures in M51 ({\it black dots}, \citealt{kennicutt07a}), azimuthal averages ({\it blue circles}) in NGC4736 and NGC5055 \citep{wong02}, NGC6946 \citep{crosthwaite07a}, and M51 \citep{schuster07a}, and global averages for starbursts ({\it open green triangles}, \citealt{kennicutt98a}), normal spirals ({\it filled green triangles}, \citealt{kennicutt98a}), and low surface brightness galaxies ({\it yellow diamonds}, \citealt{wyder09a}). The gray arrows and labels indicate schematically the dominant physical process responsible for setting the slope in each region.
}
\end{figure*}

We compare our proposed star formation law, equation (\ref{sfreq2}), to the observed relationship between star formation, atomic gas, and molecular gas in Figures \ref{sflaw1} and \ref{sflaw2}. The majority of the observations come from the THINGS sample. The full sample covers metallicities from $\log Z'=-1.22$ to $0.49$ (\citealt{walter08a}; \citetalias{krumholz08d}), but only four of the thirty-four galaxies have metallicities below $\log Z'=-1.0$, and these are all dwarfs with such low star formation rates that they contribute negligibly to the total star formation rate in the sample. Moreover, the molecular gas masses for these systems are likely to be extremely uncertain (see below). Thus we adopt $\log Z'=-1.0$ to $0.5$ as a realistic range of metallicities in the data.

The THINGS sample is observed at a resolution of $\sim 750$ pc, much larger than a single atomic-molecular complex, so we expect $c>1$. The true value of $c$ cannot be determined directly in external galaxies without higher resolution observations. A lower limit comes from the fact that the observations mix together spiral arm and inter-arm regions, and the arm-interarm density contrast is $\sim 2-4$ in galaxies observed at higher resolution \citep{nakanishi03, schuster07a}. The complexes themselves represent density peaks on top of the already-enhanced density within the arm, and in fully molecular regions clouds are observed to have surface densities higher than the mean by a factor of $\sim 2$ \citep{rosolowsky05a}. We therefore adopt $c\approx 5$, and thus we expect the data to be characterized by $\log cZ' \approx -0.3$ to $1.2$, with the four low-metallicity dwarfs lying at somewhat lower $\log cZ'$.

Our simple model recovers a number of salient features in the observations. Figure \ref{sflaw1}a shows that we recover the observational result that the \hi\ surface density reaches a maximum value, which is $\sim 10$ $\msun$ pc$^{-2}$ at Solar metallicity, and that the star formation rate does not correlate well with $\Sigma_{\rm HI}$ in resolved observations of galaxies. The RMS noise in the star formation rate surface density in the survey is $\sim 10^{-4}$ $\msun$ kpc$^{-2}$ \citep{bigiel08a}, so the apparent flattening of the contours below this value is an observational artifact. Figure \ref{sflaw1}b indicates that we recover a good approximation to the correct, nearly constant star formation rate in molecular gas at surface densities from roughly $5-100$ $\msun$ pc$^{-2}$. Combined, these two effects produce a star formation rate that increases superlinearly with total gas content below the \hi\ saturation threshold and only linearly above it (Figure \ref{sflaw1}c). Third, we recover the return to a superlinear increase of star formation rate with total gas content above $\sim 100$ $\msun$ pc$^{-2}$, produced by the increase in molecular cloud density in high-pressure environments (Figure \ref{sflaw2}).

It is worth noting that the observations are subject to significant systematic errors in both the gas surface density and in the star formation rate, and that this is likely to limit the extent of agreement between the data and any theoretical model, including this one. The main uncertainty in the gas surface density is in the factor $X_{\rm CO}$ used to convert an observed CO intensity into a molecular cloud mass. The standard assumption of a constant $X_{\rm CO}=2.0\times 10^{20}$ cm$^{-2}$ $(\mbox{K km s}^{-1})^{-1}$ is likely to be accurate to a factor of $\sim 2$ in spiral galaxy environments, where its value has been calibrated against other methods of estimating mass \citep{blitz07a}. In low metallicity dwarf galaxies, however, there is evidence that the assumption of a fixed $X_{\rm CO}$ may underestimate the true molecular mass by as much as a factor of $\sim 10$ \citep{leroy07a}, while in starburst systems it may overestimate the mass by factors of up to $\sim 5$ \citep{downes98}. The variation of $X_{\rm CO}$ in starbursts has been at least approximately accounted for in the high $\Sigmag$ systems shown in Figure \ref{sflaw2}, but the uncertainty there is probably larger than for spiral galaxies. For dwarf systems at low metallicity, on the other hand, the values of $\Sigmag$ and $\Sigma_{\rm H_2}$ in Figures \ref{sflaw1} and \ref{sflaw2} have for the most part been derived assuming a constant $X_{\rm CO}$, and thus the possible systematic underestimate of $\Sigmag$ and $\Sigma_{\rm H_2}$ has not been included. This would tend to shift points to the right in Figure \ref{sflaw1}c and Figure 2. The effect in Figure \ref{sflaw1}b will be minimal, because most of the galaxies for which this effect is significant fall below the completeness limit in any event.

Uncertainties in the star formation rates come from a combination of uncertainties in dust corrections and in the stellar initial mass function (IMF). Comparing star formation rates in the THINGS sample based on FUV plus 24 $\mu$m emission to those based on H$\alpha$, or H$\alpha$ plus 24 $\mu$m emission, suggests uncertainties below the factor of $\sim 2$ level. Comparison to the star formation rate in the Milky Way inferred from radio catalogs of H~\textsc{ii} regions suggests a slightly larger uncertainty: \citet{mckee97} infer a star formation timescale in the molecular gas of $(\sfrff/\tff)^{-1} = 300$ Myr from this technique, compared to 2 Gyr for the average of the THINGS sample. The origin of this discrepancy is unclear, but it suggests that significant caution is warranted in interpreting the star formation rates inferred from observations.

\section{Discussion}

\subsection{Relation to the Previous Work}

It is important to understand how our results here relate to previous work on star formation laws, both phenomenological and theoretical. Our new star formation law, equation (\ref{sfreq2}), is more complex than the simple \citet{kennicutt98a} powerlaw $\dot{\Sigma}_* \propto \Sigmag^{n}$, with $n= 1.4\pm 0.15$, and the related observational correlation between the CO and infrared luminosities of galaxies (assumed to be proxies for the molecular gas mass and the star formation rate, respectively) reported by \citet{greve05} and \citet{riechers06a}. Such increased complexity is demanded by improving observations, since the data shown in Figures \ref{sflaw1} and \ref{sflaw2} clearly cannot be fit by single powerlaws. There are very distinct regions where the relationship between $\dot{\Sigma}_*$ and $\Sigmag$ is steep, flattens, and then steepens again. That said, our model is in fact consistent with the data set on which the \citet{kennicutt98a} law is based -- these are shown by the green open and filled triangles in Figure \ref{sflaw2} -- and these data in turn fall within the contours of the THINGS observations. Thus there is no inconsistency in the data themselves. 

Why then is it possible to fit the data of the \citet{kennicutt98a} sample with a single powerlaw? The answer comes partly from the fact that the data are averaged over entire galaxies, which introduces significant scatter compared to the more recent data that is resolved to sub-kpc scales. The primary reason for the single powerlaw Kennicutt fit with an index of $n\approx 1.4$, however, is that most of the dynamic range in $\Sigmag$ that gives rise to the index of $1.4$ comes from galaxies with $\Sigmag \gtsim 100$ $\msun$ pc$^{-2}$. The same is true of the observed correlation between CO and IR luminosities in galaxies, where most of the dynamic range in the sample comes from starbursts with large surface densities and star formation rates. In this regime our predicted law, equation (\ref{sfreq2}), also reduces to a simple powerlaw $\dot{\Sigma}_*\propto \Sigmag^{1.33}$; the index $1.33$ is well within the error bars in Kennicutt's fit. Conversely, a fit to only the normal galaxies in Kennicutt's sample produces a much steeper best-fit index of $n=2.47 \pm 0.39$, consistent with the steeper slope we predict for normal galaxies due to the dependence of the H$_2$ fraction on $\Sigmag$. Thus the classic single powerlaw star formation law is in part an artifact of fitting a single powerlaw between normal galaxies and starbursts, a point also made by \citet{gao04a} and \citet{wyder09a}.

On the theoretical side, the classical explanation for the observed $\dot{\Sigma}_*\propto \Sigmag^n$ and $L_{\rm IR}\propto L_{\rm CO}^{m}$ correlations with $n\approx m\approx 1.5$ is that the star formation rate should be proportional to the gas mass divided by the free-fall time \citep{madore77, elmegreen94b, kennicutt98a, elmegreen02}. The latter varies as density to the $-1/2$ power, so in a parcel of gas of density $\rho$ the star formation rate per unit volume varies as $\rho^{3/2}$. The model we propose here is entirely consistent with this basic picture, and with earlier explanations for correlations between star formation rates and gas masses and surface densities by \citetalias{krumholz05c} and \citet{krumholz07g}, despite the fact that we obtain a nearly linear correlation between $\dot{\Sigma}_*$ and $\Sigma_{\rm H_2}$ for $\Sigmag \ltsim 100$ $\msun$ pc $^{-2}$.

The reason we can obtain a linear correlation between $\dot{\Sigma}_*$ and $\Sigma_{\rm H_2}$ is that, in any theory in which the star formation timescale is proportional to the free-fall time, the star formation rate per unit area will vary as $\dot{\Sigma}_* \propto \Sigma_{\rm H_2}/t_{\rm ff} \propto \Sigma_{\rm H_2} \rho^{1/2}$. Similarly that the total star formation rate in a galaxy will vary as $\dot{M}_*\propto M_{\rm H_2} \rho^{1/2}$, where $M_{\rm H_2}$ is the total mass of molecular gas. The indices $n$ and $m$ for the areal and total star formation laws therefore depend on how $\rho$ varies with $\Sigma_{\rm H_2}$ and $M_{\rm H_2}$, respectively. The most common assumption in previous work has been to adopt $\rho\propto \Sigmag$ (e.g.\ \citealt{elmegreen02}) and $\rho\propto M_{\rm H_2}$, but this assumption is independent of the basic idea that the star formation rate varies inversely with the free-fall time. Indeed, if the characteristic densities of star-forming clouds do not vary from galaxy to galaxy, as is approximately (though not exactly) the case in our model for galaxies with $\Sigmag < 85$ $\msun$, then we expect a linear relationship between $\dot{\Sigma}_*$ and $\Sigma_{\rm H_2}$, even though the underlying volumetric star formation law is nonlinear. \citet{krumholz07g} obtain a linear relationship between the luminosities of galaxies in the infrared and in HCN($1\rightarrow 0$) emission for much the same reason. They point out that the high critical density of the HCN line ensures that it traces gas at similar densities in every galaxy, at least up to the most extreme starbursts. Since $\rho$ is constant, the star formation law is linear. In our model here $\rho$ is roughly constant in galaxies with $\Sigmag\ < 85$ $\msun$ pc$^{-2}$ for a different reason -- GMC densities are unaffected by galactic environment because GMC pressures greatly exceed mean ISM pressures -- but the end result, a linear star formation law, is the same. Conversely, a scaling law $\dot{\Sigma}_* \propto \Sigma_{\rm H_2}^{3/2}$ or $L_{\rm IR} \propto L_{\rm CO}^{3/2}$ is expected when the characteristic density of the molecular gas scales close to linearly with its surface density or total mass. In our model such a scaling appears in the starburst regime because ISM pressure becomes important in setting GMC internal densities. 

\subsection{Predictions for Future Observations}

Our model makes distinct predictions that can be tested against future observations. One obvious prediction is the upward kink in the relationship between $\dot{\Sigma}_*$ and $\Sigma_{\rm H_2}$, or $\dot{\Sigma}_*$ and $\Sigmag$, seen at $\Sigma_{\rm H_2} \approx \Sigmag\approx 100$ $\msun$ pc$^{-2}$ (Figures 1a, 1b, and 2). This kink in our model is caused by the transition between molecular cloud surface densities being determined by internal regulation and being determined by external pressure. A spatially resolved survey such as THINGS targeting the nearest circumnuclear starburst galaxies such as M64, which reach surface densities in the range $\sim 50-1000$ $\msun$ pc$^{-2}$ \citep{rosolowsky05a}, should reveal this kink quite clearly, perhaps even within a single galaxy. The one caveat is that locating the kink depends on being able to measure molecular surface densities accurately, which in turn depends on our knowledge of the X factor used to convert CO intensity to surface density. Uncertainties in its value translate directly into uncertainties in the location of the kink in the star formation law.

A second testable prediction comes from our predicted metallicity-dependence of the relationship between $\dot{\Sigma}_*$ and $\Sigmag$ (Figure 1c). Our model curves span the range of metallicities covered by the THINGS sample, and, as of the date of this paper's submission, a version of the data shown in Figure 1 binned by metallicity was not available. However, such metallicity binning should reveal that low metallicity galaxies have systematically lower star formation rates at fixed $\Sigmag$ for galaxies when $\Sigmag \ltsim 10$ $\msun$ pc$^{-2}$. Because there is degeneracy between the metallicity $Z'$ and the clumping factor $c$, ideally this prediction would be tested with higher resolution observations that reduce or remove the need to adopt a clumping factor to account for the structure of the gas on scales unresolved by the observations. Surveying a large sample of entire galactic disks with $\sim 100$ pc resolution, so that $c\approx 1$, is probably prohibitively expensive with the current generation of telescopes. However, even a random sample of sightlines through galaxies at varying $\Sigmag$ and metallicity, with $\sim 100$ pc resolution, should be sufficient to test for a metallcity-dependent correlation between star formation rate and gas surface density such as the one we predict.

\section{Summary}

We have shown that the observed relationship between the star formation rate and the atomic and molecular content of galaxies can be explained by a simple model, whose elements are summarized by the regions labelled in Figure \ref{sflaw2}. First, self-shielding of hydrogen determines the amount of gas in molecular form. This imposes a characteristic gas surface density of $\sim 10/cZ'$ $\msun$ pc$^{-2}$ for the transition from atomic to molecular, where $c$ is the factor by which the gas surface density is increased due to clumping unresolved by the observations and $Z'$ is the metallicity relative to solar. Second, once molecules do form, molecular clouds reach a surface density of roughly $85$ $\msun$ pc$^{-2}$ independent of galactic environment. This behavior can be understood as arising from the fact that molecular clouds are overpressured relative to their surroundings, so they must be regulated by internal processes, most likely H~\textsc{ii} regions \citep{matzner02, krumholz06d}, that do not depend on metallicity or other large-scale galaxy properties. The constant surface density imposes a roughly constant volume density and free-fall time on all molecular gas. The exception to this is galaxies where the mean galactic surface density is $\gtsim 100$ $\msun$ pc$^{-2}$, in which the ambient pressure is high enough to force GMC densities to rise along with galactic surface density in order to keep the clouds in pressure balance. Third, once formed molecular clouds convert themselves into stars at a nearly universal rate of $\sim 1\%$ of the mass per free-fall time as a result of turbulent regulation. Together these effects produce a total gas star formation law that is superlinear at low galactic column density (due to increasing molecular fraction), linear or slightly sub-linear at intermediate column (due to the invariance of molecular cloud surface densities and the weak dependence of GMC masses on galactic properties), and superlinear again at high column (due to the breakdown of this invariance at high galactic pressures).

It is worth noting that our model does not make any explicit reference to galactic-scale processes such as spiral shocks, gravitational instability, supernova feedback, or cloud-cloud collisions. In a sense all of this physics is ``upstream" of our theory: processes such as these are almost certainly responsible for determining the distribution of gas surface density within a galaxy. Our model addresses the next step of how, once large-scale processes assemble the gas, some fraction of it forms molecular clouds and then turns into stars. We have therefore separated the problem of star formation into two parts, and provided a tentative solution for one of them: galactic-scale processes determine $\Sigmag$, but the physics responsible for determining the star formation rate thereafter is purely local, and can be understood without reference to galactic-scale behavior. Our model shows that a much of the recent observational work on star formation can be understood in terms of a simple model for that local process.

\acknowledgements We thank F.~Bigiel for providing a copy of his data. We thank the anonymous referee for helpful comments. Support for this work was provided by the Alfred P.\ Sloan Foundation (through a Sloan Research Fellowship to MRK), by NASA, as part of the Spitzer Theoretical Research Program, through a contract issued by the JPL (MRK), and by the National Science Foundation through grants AST-0807739 (to MRK), AST-0606831 (to CFM), and PHY05-51164 (to the Kavli Institute for Theoretical Physics).

\end{document}